\documentclass[aps,prd,twocolumn,showpacs,floatfix,preprintnumbers,nofootinbib,superscriptaddress]{revtex4}
\usepackage{mathtools}
\usepackage{graphicx}
\usepackage{color}
\usepackage{natbib}
\usepackage{multirow}
\usepackage{amsmath}
\usepackage[amssymb]{SIunits}

\newcommand{\itm}{\item[\textbullet]}
\begin{document}

\title{Nonuniversal BBN bounds on electromagnetically decaying particles}
\author{Vivian Poulin}
\author{Pasquale Dario Serpico}
\affiliation{LAPTh, Univ. de Savoie, CNRS, B.P.110, Annecy-le-Vieux F-74941, France}

\date{\today}

\preprint{LAPTH-013/15}

\begin{abstract}
In Poulin and Serpico [Phys. Rev. Lett. 114, 091101 (2015)] we have recently argued that when the energy of a photon injected in the primordial plasma falls below the pair production threshold, the universality of the nonthermal photon spectrum from  the standard theory of 
electromagnetic cascades onto a photon background breaks down. We showed that this could reopen or widen
the parameter space for an exotic solution to the ``lithium problem.'' Here we discuss another application, namely the impact that this has on nonthermal big bang nucleosynthesis  constraints from  ${}^4\textrm{He}$, ${}^3\textrm{He}$ and ${}^2\textrm{H}$, using the parametric example of monochromatic photon injection of different energies. Typically,  we find tighter bounds than those existing in the literature, up to more than 1 order of magnitude.  As a consequence of the nonuniversality of the spectrum, the energy dependence of the photodissociation cross sections is important. We also compare the constraints obtained with current level and future reach of cosmic microwave background spectral distortion bounds.
 
\end{abstract}

\maketitle
\section{Introduction}

Big bang nucleosynthesis (BBN) has been used for decades as a very powerful tool to constrain exotic particle physics (for reviews, see for instance Refs. ~\cite{Iocco:2008va,Pospelov:2010hj}.) In particular, metastable particles populating the plasma in the early Universe  could induce a nonthermal BBN phase via their decay products. Both hadronic and electromagnetic cascades typically contribute to these phenomena, although the former are much more model dependent. On the contrary, electromagnetic cascades are known to lead to a quasiuniversal  $\gamma$-spectrum, only dependent on the overall energy injected and the injection time: a monotonically decreasing, broken power-law  (see e.g. Chapter VIII in Ref.~\cite{Ginzburg:1990sk} for a basic derivation).
Recently, we pointed out that in a particular regime the commonly used universality of the spectrum
achieved by photons as a result of electromagnetic cascades is violated~\cite{PaperI}. 
This corresponds to the situation when the photons injected at energy  $E_\gamma$ are not sufficiently energetic to induce pairs onto the background photons at temperature $T$, and can be translated in the condition   $E_\gamma\lesssim 10\, T_{\rm keV}^{-1}\,$MeV (we use natural units with $c=k_B=1$). 
For $T$ of order ${\cal O}$(keV) down to ${\cal O}$(eV) characteristic of the period between the end of BBN to the formation of the cosmic microwave background (CMB), these energies are typically higher than the photodisintegration thresholds of light nuclei, denoted by  $E_{\rm th}$. 
As a result, the injection of photons of which the energy falls in the couple of decades from a few MeV to a few hundreds MeV may have an impact different than the one estimated with the universal spectra, given by
\begin{equation} \label{eq:spectrum}
\frac{dN_\gamma }{dE_\gamma} = \left\{ \begin{array}{cl}
& K_0\left(\frac{E_\gamma}{\epsilon_X}\right)^{-3/2} \textrm{for } E_\gamma < \epsilon_X\,, \\
& K_0 \left(\frac{E_\gamma}{\epsilon_X}\right)^{-2} \:\:\;\textrm{for $\epsilon_X \leq E_\gamma \leq \epsilon_c$}\,, \\
& 0 \:\:\:\:\:\:\:\:\:\:\:\:\:\:\:\:\:\:\:\:\:\:\:\textrm{ for } E > \epsilon_c\,.
\end{array}\right.
\end{equation} 
In the above expression, $K_0 = E_0\epsilon_X^{-2}[2+\ln(\epsilon_c/\epsilon_X)]^{-1}$ is a normalization constant that enforces the condition that the total energy is equal to the injected electromagnetic energy, $E_0$;
 the characteristic energy $\epsilon_c= m_e^2/\epsilon_\gamma^{\rm max}$ denotes the effective
threshold for pair production, $\epsilon_\gamma^{\rm max}$ being the highest energy of the photon background onto which
pairs can be effectively created;  and $\epsilon_X \lesssim \epsilon_c/3 $ is the maximum energy of up-scattered inverse Compton (IC) photons (See Refs. ~\cite{Ellis:1990nb, Kawasaki:1994sc,Protheroe:1994dt} for Monte Carlo studies leading to further justification of these parameters.)

In our previous publication~\cite{PaperI}, we illustrated how this may reopen the possibility of purely electromagnetic new physics solution to the so-called lithium problem, but we anticipated that other domains may be affected. Here we outline the impact on the constraints in the abundance vs lifetime plane for unstable early Universe relics, decaying electromagnetically, and derived from the deuterium, $^4$He and $^3$He measurements. Our main conclusion is that the bounds are non universal and that they may be significantly more stringent than commonly thought.
In the following, we will compare the constraints obtained from different elements in the hypothesis of the universal spectrum
with the actual constraints obtained for monochromatic photon injections at different energies, below the pair production
threshold $\epsilon_c$.  This parametrization is used solely for the sake of clarity; the differences would persist for any spectrum (either primary photons or secondary due e.g. to upscattering of background photons via the IC by energetic $e^\pm$) injected below the critical energy.

This article is structured as follows. In Sec.~\ref{spectrum}, we describe the features of the electromagnetic (e.m.) cascades and the breakdown of the universal nonthermal spectrum, as well as our method to solve the relevant Boltzmann equations. In Sec.~\ref{form}, we describe the nonthermal nucleosynthesis formalism and the observational constraints
being used in the following.  In Sec.~\ref{fromCMB} we review the constraints coming from the CMB, notably from its spectral distortions, to which we will compare the BBN ones.
Our results are reported in Sec.~\ref{results}. Finally, Sec.~\ref{concl} contains a discussion with
our conclusions.

\section{E.m. cascades and  breakdown of universal nonthermal spectrum} \label{spectrum}
In general, in order to compute the nonthermal photon spectrum which can photodisintegrate nuclei, one has to follow
the coupled equations of both photon and electron-positron populations. For the problem at hand, however, where we limit ourselves to inject photons {\it incapable} of pair production,  it is a good first approximation to ignore the nonthermal electrons; while the injected photons will in general Compton scatter and produce them, a further process, typically IC onto the photon background, is needed to channel back part of their energy in the photon channel. The energy of these photons is significantly lower than the injected photon one: whenever they are reinjected below nuclear photodissociation thresholds they are actually lost for nonthermal nucleosynthesis; otherwise they would contribute to {\it strengthening} the bounds, although only by a few percent, for the cases discussed below. 
For simplicity, let us also start by assuming that  all photon interactions are {\it destructive}; i.e. photons are not downscattered to a lower energy.
 Within this approximation, the Boltzmann equation describing the nonthermal photon distribution function $f_\gamma$ reads
\begin{equation}\label{eq:BoltzmannPh}
\frac{\partial f_\gamma(E_\gamma)}{\partial t} = -\Gamma_\gamma(E_\gamma ,T(t))f_\gamma(E_\gamma,T(t))+{\cal S}(E_\gamma,t)\,,
\end{equation}
where ${\cal S}(E_\gamma,t)$ is the source injection term, $\Gamma_\gamma$ is the total interaction rate, and we neglected the Hubble expansion rate since interaction rates are much faster and rapidly drive $f_\gamma$ to a quasistatic equilibrium, $\frac{\partial f_\gamma(\epsilon_\gamma)}{\partial t} =0$.
Thus, we simply have
\begin{equation}\label{eq:Spectre2}
f^{\textrm{S}}_\gamma(E_\gamma,t)= \frac{{\cal S}(E_\gamma,t)}{\Gamma_\gamma(E_\gamma,t)}\,,
\end{equation}
where the term ${\cal S}$  for an exponentially decaying species with lifetime $\tau_X$ and density $n_X(t)$, of which the total e.m. energy injected per particle is $E_0$, can be written as 
\begin{equation}\label{eq:SSpectre}
{\cal S}(E_\gamma,t)  =  \frac{n_\gamma^0\zeta_X(1+z(t))^3\,e^{-t/\tau_X}}{E_0\tau_X}\,p_\gamma(E_\gamma,t) \,,
\end{equation}
with $z(t)$ being the redshift at time $t$ and the energy parameter $\zeta_X$ (conventionally used in the literature) is simply defined in terms of the initial comoving density of the $X$ particle $n_X^0$ and the actual one of the CMB, $n_\gamma^0$, via $n_X^0=n_\gamma^0\zeta_X/E_0$.  We shall use as a reference spectrum the one for a two body decay $X\to \gamma\,U$ leading to a single monochromatic line of energy $E_0$,  corresponding to $p_\gamma(E_\gamma)=\delta(E_\gamma-E_0)$. If the unspecified particle $U$  is (quasi)massless, like a neutrino, one has $E_0=m_X/2$, where $m_X$ is the mass of the decaying particle. Note that here, we will be interested in masses $m_X$ between a few and ${\cal O}(100)$ MeV, and at temperatures of order few keV or lower, hence the thermal broadening is negligible, and a Dirac delta spectrum as the one above is appropriate.

We calculate $\Gamma_\gamma$ by summing the rates of processes that degrade the injection spectrum, namely:
\begin{enumerate}
\itm Scattering off thermal background photons, $\gamma_{th}$: $ \gamma +  \gamma_{th} \rightarrow \gamma +\gamma   $. \\
This has been studied in Ref. \cite{Svensson:1990}, and the scattering rate of a $\gamma$-ray with energy $E_\gamma$ over a blackbody with temperature $T$ is given by 
\begin{equation}
\Gamma_{\gamma\gamma}= -0.1513\alpha^4 m_e\bigg(\frac{E_\gamma}{m_e}\bigg)^3\bigg(\frac{T}{m_e}\bigg)^6\,.
\end{equation}
\itm Bethe-Heitler pair creation : $ \gamma +  N \rightarrow e^\pm +N  $. \\
The cross section for this process is given by \cite{Jedamzik:2006xz}  
\begin{equation}
\sigma_{\textrm{BH}} \simeq \frac{3}{8}\frac{\alpha}{\pi}\sigma_{\textrm{Th}}\left(\frac{28}{9}\ln\left(\frac{2E_\gamma}{m_e}\right)-\frac{218}{27}\right)Z^2\,.
\end{equation}
\itm Compton scattering over a thermal electron :
$\gamma +  e^{\pm}_{th} \rightarrow \gamma + e^{\pm}$. 
For the temperature of interest of $\mathcal{O}$(keV), one can consider electrons to be at rest. In this case, we have \cite{Kawasaki:1994sc}
\begin{eqnarray}
\Gamma_{\textrm{CS}} & =  & \bar{n}_{e}\frac{3\sigma_{\textrm{Th}}}{4x}\times\\
&  & \left[\left(1-\frac{4}{x}-\frac{8}{x^2}\right)\ln\left(1+x\right)+\frac{1}{2}+\frac{8}{x}-\frac{1}{2(1+x)^2}\right]\,,\nonumber
\end{eqnarray}
where $x=\frac{2E_{\gamma}}{m_e}$ and  $\bar{n}_e$ is the number density of background electrons and positrons.
\end{enumerate}

In reality, not all scattered  photons will be ``lost'':  even ignoring the energy transferred to $e^-$ and $e^+$, Compton scattering and $\gamma\gamma$ scattering still leave lower-energy photons in the final state. This effect can be accounted for by replacing  the rhs of Eq.~(\ref{eq:BoltzmannPh}) by the following term:
\begin{equation}\label{eq:reinjection}
{\cal S}(E_\gamma,t)\to {\cal S}(E_\gamma,t) + \int_{E_\gamma}^\infty dx K_{\gamma}(E_\gamma,x, t)f_\gamma(x\,,t)\,.
\end{equation}

\begin{figure}
\centering
\includegraphics[width=0.49\textwidth]{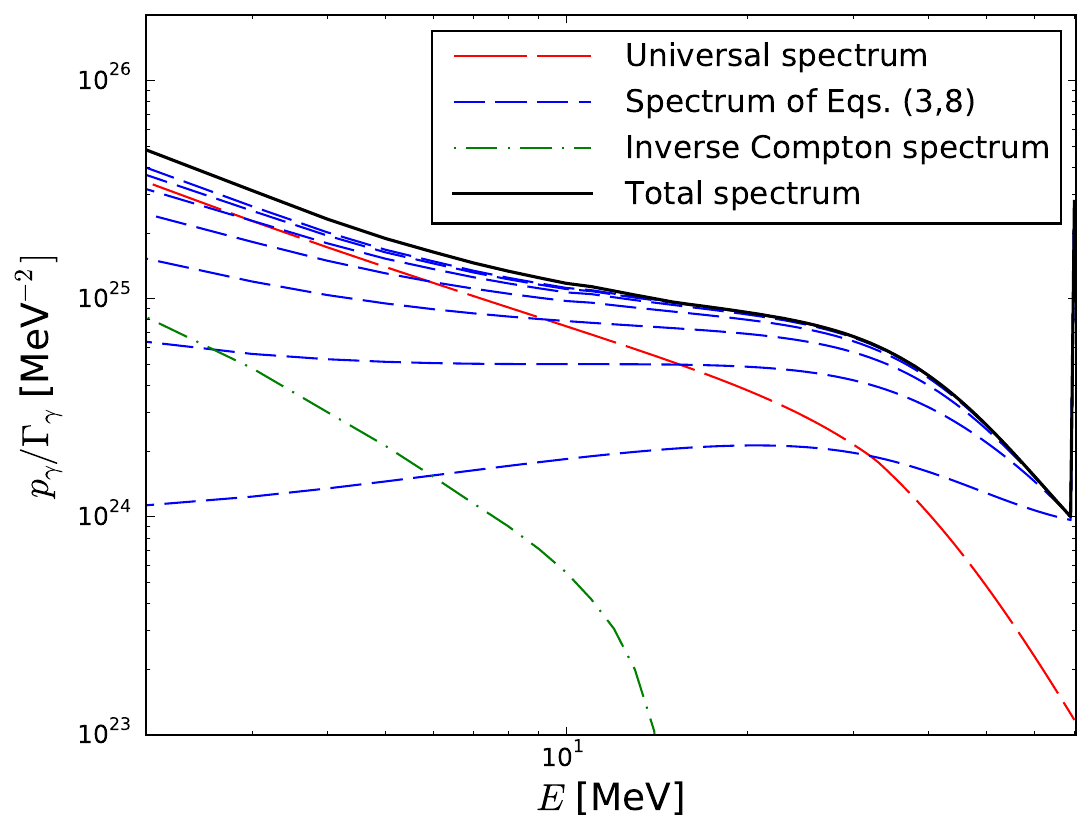}
\caption{\label{fig:iteration} Spectrum computed in this article (solid black line) 
compared with approximated one used in the literature (long-dashed red line), for the case $E_0 = 70\,$MeV at $T=100\,$eV. The short-dashed, blue lines show the contribution from the photon population as computed in our iterative treatment, with the number of iterations increasing from 1 to 7
from bottom to top. The dot-dashed curve is the estimated contribution to the photon spectrum from the nonthermal population of electrons excited by the energy loss
channels of the photons.}
\end{figure}

The additional term of which the kernel is $K$ accounts for scattered photons and is obtained by summing the differential rates for the $\gamma\gamma$ scattering off background photons and the Compton scattering over thermal electrons, respectively given by~ \cite{Svensson:1990}
\begin{eqnarray}
\frac{d\Gamma_{\gamma\gamma}(E_\gamma,E_\gamma')}{dE_\gamma'}& & = \frac{1112}{10125\pi}\alpha^2r_\epsilon^2m_e^{-6}\times\frac{8\pi^4T^6}{63}\times\nonumber\\
& &  E_
\gamma'^2\left[1-\frac{E_\gamma}{E_\gamma'}+\left(\frac{E_\gamma}{E_\gamma'}\right)^2\right]^2\,, \\
\frac{d\Gamma_{\textrm{CS}}(E_\gamma,E_\gamma')}{dE_\gamma'}&  & =  \pi r_e^2\bar{n}_e\frac{m_e}{E^{'2}_{\gamma}}\times\\
& & \left[\frac{E_{\gamma}'}{E_{\gamma}}+\frac{E_{\gamma}}{E_{\gamma}'}+\left(\frac{m_e}{E_{\gamma'}} - \frac{m_e}{E_{\gamma}} -1\right)^2-1\right]\,.\nonumber
\end{eqnarray}
The integral in Eq.~(\ref{eq:reinjection}) now depends on $f_\gamma$. 
We numerically solve this Boltzmann equation  using an iterative method: we start from the Dirac distribution and the algebraic solution of Eqs.~(\ref{eq:Spectre2}) and (\ref{eq:SSpectre}), plug in the result thus obtained in Eq.~(\ref{eq:reinjection})  to estimate the new ``effective'' source term, and proceed. Note that the zeroth-order solution of Eqs.~(\ref{eq:Spectre2}) and (\ref{eq:SSpectre}) is exact at the end point $E_\gamma=E_0$, with further iterations essentially improving the description at lower and lower energies. We stop iterating when the resulting improvement on the constraints is smaller than 3\%. Figure~\ref{fig:iteration} shows the resulting spectrum proportional to $f_\gamma$ according to the prefactor of  Eq.~(\ref{eq:SSpectre})] for an injected monochromatic photon of 70 MeV at the temperature $T = 100\,$eV in the commonly used universal spectrum approximation (long-dashed red line) and for the actual solution of the Boltzmann equation, as a function of the iteration (short-dashed blue lines). For this example, one can estimate $\epsilon_c\simeq 100\,$MeV and $\epsilon_X\simeq 30\,$MeV.  Two features are clearly visible: i) the universal spectrum grossly fails for $E_\gamma\gtrsim \epsilon_X$, as expected, since it imposes an artificial suppression; ii) the exact solution is significantly harder at intermediate energies, but attains the same slope as the universal spectrum at low energies. However, the low-energy normalization is altered, since the universal spectrum is unphysical in pushing too many photons to low energies (below nuclear thresholds).

  Although neglected so far, an analogous treatment can be applied to the nonthermal electron distribution $f_e$: the source term will be given by the Compton scattering and Bethe-Heitler processes of nonthermal photons, and the ``loss term'' into the photon channel essentially by IC scattering. The latter will in turn correspond to a new source term in Eq.~(\ref{eq:reinjection}), the impact on the photon spectrum is reported in Figure~\ref{fig:iteration} with a dot-dashed green line.
It is clear that, unless the injected photon energies are too high, this only brings a modest correction to the {\it low-energy} tail of the spectrum, with the expected improvement in the constraints being even less prominent. The iterative solution technique adopted above would still perform correctly, although a detailed evaluation would render the
calculation unnecessarily lengthy, and will not be pursued further here. We checked in fact that, for the cases discussed in the following, four iterations for the photon spectrum are enough to obtain bounds accurate at the 10\% level (and often better) and always on the conservative side.

Since the critical energy for pair production is a dynamical quantity that increases at later times due to the cooling of the Universe, it may happen that the primary photons energy $E_0$ is above threshold for pair production at early times and below it at late times (we do take into account that the decay is not instantaneous). In general, at each time we will compare 
$E_0$ with $\epsilon_c$ and use the universal spectrum when $E_0>\epsilon_c$ or the monochromatic spectrum with the complete expression for $\cal{ S}$ when $E_0<\epsilon_c$. This gives always a qualitatively correct solution, albeit it is somewhat approximate when $E_0\sim \epsilon_c$. Since this is realized only in a very narrow interval of time, however, the final results are also quantitatively robust, barring artificial "fine-tuned" results in a specific region of the parameter space.

\section{ nonthermal nucleosynthesis} \label{form}
\subsection{Review of the formalism}
At temperatures of few keV or lower, the standard BBN is over, and the additional nucleosynthesis can be simply dealt with  as a postprocessing of the abundances computed in the standard scenario, for which we use the 
 input values from  {\sf Parthenope}~\cite{Pisanti:2007hk},  with the updated value of $\Omega_b$ coming from Ref. \cite{Planck:2015xua}.

As long as the amount of injected energy is small compared to the density of background CMB photons, one can neglect its impact on the expansion history. Thus, the nonthermal nucleosynthesis due to electromagnetic cascades can be described by a system of coupled differential equations of the type
\begin{eqnarray}
\frac{dY_A}{dt} & = & \sum_{{T}}Y_T\int_0^\infty dE_\gamma f_\gamma(E_\gamma,t) \sigma_{\gamma+T\rightarrow A}(E_\gamma) \nonumber \\
& - & Y_A  \sum_{{P}}\int_0^\infty dE_\gamma f_\gamma(E_\gamma,t) \sigma_{\gamma+A\rightarrow P}(E_\gamma)\,,\label{nonthBBN}
\end{eqnarray} 
where: $Y_A \equiv n_A/n_b$ is the ratio of the number density of the nucleus $A$ to the total baryon number density $n_b$ (this 
factors out the trivial evolution due to the expansion of the Universe); and $\sigma_{\gamma+T\rightarrow A}$ is the photodissociation cross section
of the nucleus $T$ into the nucleus $A$, i.e. the production channel for $A$;  $\sigma_{\gamma+A\rightarrow P}$ is the analogous destruction channel. Both cross sections
are actually vanishing below the corresponding thresholds. In general one also needs to follow secondary reactions of the nuclear byproducts of the photodissociation,
which can spallate on or fuse with background thermalized target nuclei, but none of that is relevant for the problem at hand. According to Ref.~\cite{Cyburt:2002uv}, the only signification secondary production is that of ${}^6$Li. Despite extensive work
in the past, the current observational status of ${}^6\textrm{Li}$ as a reliable nuclide for cosmological constraints is doubtful, given than most claimed detections have not been robustly confirmed, and a handful of cases are insufficient to start talking of a ``cosmological'' detection, see~\cite{Iocco:2012vg}. We shall thus conservatively ignore this nuclide and the secondary reactions in the following. 

With standard manipulations, namely by transforming Eq.~(\ref{nonthBBN}) into redshift space, defining $H(z)=H_{r}^0(1+z)^2$ as appropriate for a Universe dominated by radiation (with $H_{r}^0 \equiv H_{0}\sqrt{\Omega^{0}_{r}}$,  $H_0$ and $\Omega^{0}_{r}$ being the present Hubble expansion rate and fractional radiation energy density, respectively), one arrives at
\begin{eqnarray}\label{eq:AbondNy}
\frac{dY_A}{dz} & = & \frac{-1}{H_r^0(z+1)^{3}} \nonumber \\ & \times & \bigg[\sum_{{T}}Y_T\int_0^\infty dE_\gamma f_\gamma(E_\gamma,z) \sigma_{\gamma+T\rightarrow A}(E_\gamma) \nonumber \\
& - & Y_A  \sum_{{P}}\int_0^\infty dE_\gamma f_\gamma(E_\gamma,z) \sigma_{\gamma+A\rightarrow P}(E_\gamma)\bigg]\,,
\end{eqnarray}
which is solved numerically for the cases of interest.

\subsection{Light element abundances}
Among light elements, we can broadly speak of an agreement of standard BBN predictions with observations for the case of ${}^4\textrm{He}$, ${}^3\textrm{He}$, and ${}^2\textrm{H}$, while at
face value the ${}^7\textrm{Li}$ yield is overpredicted by a factor $\sim 3$ with respect to observations. Since the interpretation of ${}^7\textrm{Li}$ observations in terms of a primordial yield is still a subject of debate, see Refs.~\cite{Iocco:2008va,Pospelov:2010hj,Iocco:2012vg}, one can consider two possibilities: either the observed values are not representative of the cosmological production mechanism, in which case
it would be meaningless to derive constraints based on those observations, or alternatively, modifications to the standard BBN scenario, including electromagnetic cascades, {\it could} reconcile the envelope of ${}^7\textrm{Li}$ observed values with a primordial origin. In our previous paper~\cite{PaperI}, to which we address for further details, we discussed the latter possibility. In the following, we will adopt the former, more conservative option, and hence we will not use ${}^7\textrm{Li}$ for constraints on e.m. cascades. 

For the observationally imposed limits, we use the following: for ${}^4\textrm{He}$, which can only be destroyed by nonthermal BBN, we just impose the  2-$\sigma$ lower limit on the mass fraction $Y_p>0.2368$ from Ref.\cite{Aver:2011bw}.
For ${}^2\textrm{H}$ we adopt the 2-$\sigma$ limit $2.56\times10^{-5} < {}^2\textrm{H}/\textrm{H} <  3.48\times10^{-5}$  from Ref. \cite{Olive:2012xf}; similar results would follow by adopting the combination value compiled in Ref.~\cite{Iocco:2008va},  namely
$2.45\times10^{-5} < {}^2\textrm{H}/\textrm{H} <  3.31\times10^{-5}$, which is also closer to the results of Ref.~\cite{Pettini:2008mq}; our interval also overlaps with the recent determination in Ref.~\cite{Riemer-Sorensen:2014aoa}.
For 
${}^3\textrm{He}$ we impose no observational lower limit, but the 2-$\sigma$ upper limit from  \cite{Bania:2002yj} ${}^3\textrm{He}/\textrm{H}<1.5\times10^{-5}$. 
It is worth noting that, had we used some alternative recent determinations such as~\cite{Izotov:2014fga} for ${}^4\textrm{He}$ or~\cite{Cooke:2013cba} for ${}^2\textrm{H}$, some mild tension with the standard BBN predictions for the value of $\eta$ recently reported, e.g., by Planck would have appeared. These discrepancies are much smaller than the one affecting ${}^7\textrm{Li}$, and could be easily
accommodated  with a more conservative error attribution:  for ${}^4\textrm{He}$ this is the conclusion supported, e.g., in Ref.~\cite{Olive:2012xf} or the recent Ref.~\cite{Cyburt:2015mya},
essentially consistent with the value we quoted above; for ${}^2\textrm{H}$ it is also a possibility suggested by the slightly anomalous dispersion of several measurements around the mean (see
e.g. the discussion in Ref.~\cite{Iocco:2008va}). Alternative possibilities to reduce the tension with one or several of these determinations include a slightly different value of $\eta$ between the BBN epoch and
the CMB one, the addition of exotic phenomena such as cascades, and possibly others. In the following we shall adopt a similar attitude to the one adopted before for  ${}^7\textrm{Li}$ and consider conservatively the 
more generous observational ranges reported above. This is also justified to ease the comparison with earlier analyses of cascade nucleosynthesis bounds, which used similar ranges. Our main emphasis
here is in fact to gauge the impact of a more correct treatment of electromagnetic cascades, rather than deriving the most aggressive bounds achievable.
Needless to say, should more precise observational values be confirmed in future studies, if in agreement with standard BBN expectations, it would be worth it to derive updated stringent bounds; if
not,  it would be interesting to rediscuss possible explanations in the context for instance of cascade nucleosynthesis, as we did for ${}^7\textrm{Li}$ in Ref.~\cite{PaperI}.  

For the current application, the network of reactions used is reported in Fig.~\ref{fig:Cross_Sections} and follows the parametrization in the appendix of \cite{Cyburt:2002uv}. [Actually, the reaction ${}^4\textrm{He}(\gamma,{}^2\textrm{H}){}^2\textrm{H}$ is significantly suppressed with respect to the others and thus not shown in the figure but is included in our numerical treatment.] Note that all cross section share the same qualitative features: they rise fast just above threshold, go through a peak  (the so-called giant dipole resonance), eventually showing  a decreasing tail at higher energies. We shall compare the bounds thus obtained with the ones coming from CMB spectral distortions and entropy production,
briefly recalled in the following section.

\begin{figure}
\centering
\includegraphics[width=0.49\textwidth]{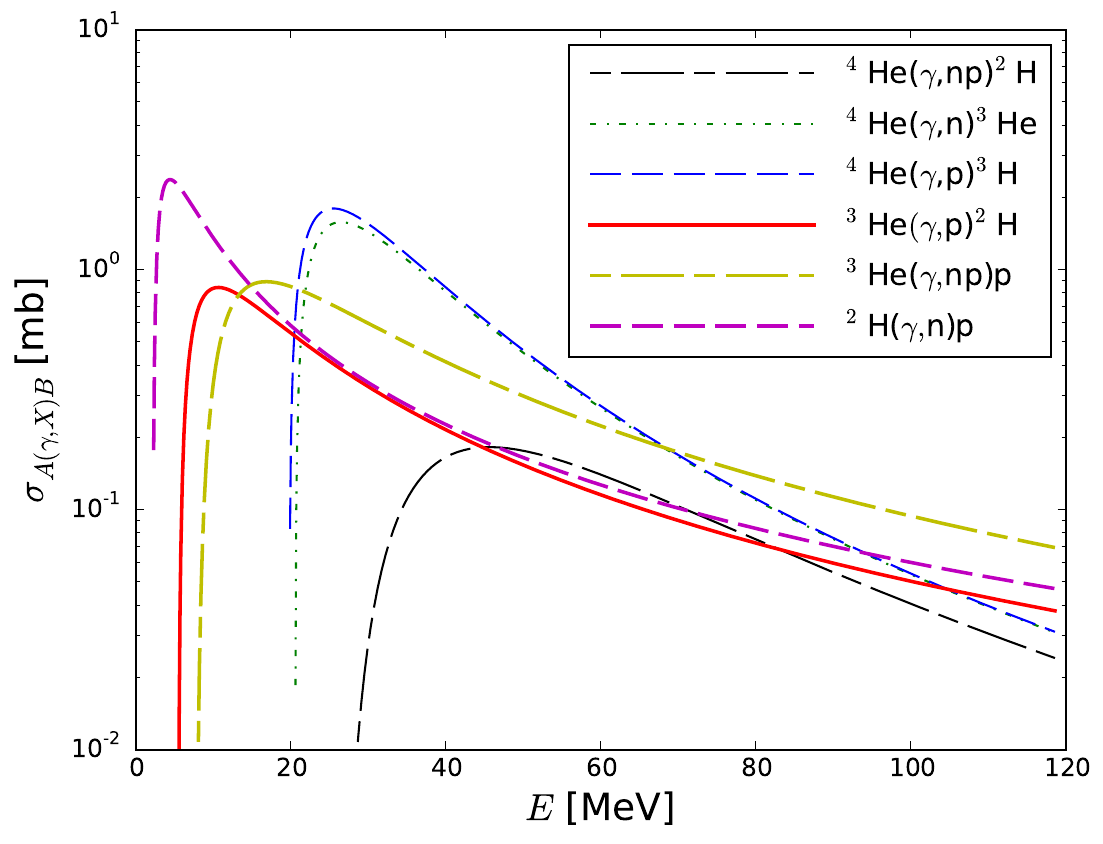}
\caption{\label{fig:Cross_Sections} Cross sections for the relevant photodisintegration processes.}
\end{figure}

\section{Constraints from the CMB}\label{fromCMB}
It is well known that a late injection of photons in the thermal bath can lead to additional measurable cosmological alterations. 

For instance, the injection of a significant amount of energy can lead to modification of the photon-baryon ratio $\eta$ or equivalently to the increase of the comoving entropy. 
Since the inferred values of $\Omega_b$ at the BBN and CMB epoch are compatible, no major entropy release could have taken place between nucleosynthesis and decoupling.
It can be shown that, in a Universe dominated by radiation and by considering that the decays have happened at $t \sim \tau$, we have for a small fractional change in entropy (see e.g. Ref.~\cite{Ishida:2014wqa}) 
\begin{equation}
\frac{\Delta S}{S} \simeq \ln \frac{S_{f}}{S_{i}} = 2.14 \times 10^{-4}\bigg(\frac{\zeta_{X\rightarrow \gamma}}{10^{-9}\textrm{ GeV}}\bigg)\bigg(\frac{\tau_{\textrm{x}}}{10^6\, \textrm{s}}\bigg)^{1/2}\,,
\end{equation}
with a slight abuse of notation since  $\zeta_{X\rightarrow\gamma}$ now has to be intended to include any electromagnetically interacting decay products, all of which contribute to modify the photon-baryon ratio. To derive a statistically sound constraint, one should combine BBN and CMB data, allowing for an  entropy increase between the two epochs. Since, as we shall see, this constraint is typically much weaker than others, such an exercise would bring us far beyond the scope of this paper adding a lengthy and unnecessary complication. We shall thus limit ourselves to illustrate the constraint that would follow by allowing  a maximal $2\%$ increase in the entropy between the two periods. This is an educated guess of the order of the bounds that one can expect, roughly corresponding to the 2-$\sigma$ error bars on $\Omega_b$ from Planck 2015 \cite{Planck:2015xua}.

Furthermore, as reviewed in detail in Ref.~\cite{Chluba:2011hw}, the spectrum of the CMB itself can also be affected through two types of deformation:  a modification of the chemical potential $\mu$ and a modification of the Compton-$y$ parameter, which is related to the energy gained by a photon after a Compton scattering. 
To first order, it is possible to distinguish the era of $\mu$ distortion from the era of Compton-$y$ distortion,
because the rate of the processes which are responsible for one type of distortions dominates at very different time. 
Basically, the $\mu$ distortion arises from rare process, implying a change in the number of photons such as $\gamma+e\rightarrow\gamma+\gamma+e$, whereas the Compton-$y$ distortions are due to the end of the equilibrium of Compton reactions, which happens much later in the history of the Universe, with a schematic  $\mu-y$ transition happening at for $z\simeq 4\times 10^5$, i.e. $\tau \simeq 5\times10^{10}$ s. For the relatively early time we focus on, the constraints come essentially from $\mu$-type distortions.
We follow here the results of Ref.~\cite{Chluba:2011hw}, which contains improvements with respect to the ones given in Ref.~\cite{Hu:1993gc}, notably for $z \lesssim 2\times 10^6$, while  Ref.~\cite{Hu:1993gc} is accurate enough at late times (see Fig. 16 in Ref.~\cite{Chluba:2011hw}). Hence, we adopt
\begin{equation}
\mu \simeq 8.01 \times 10^2 \bigg(\frac{\tau_{X}}{1\textrm{ s}}\bigg)^{1/2}\times \mathcal{J}
\times \bigg(\frac{\zeta_{X\rightarrow \gamma}}{1\textrm{ GeV}}\bigg)\,,
\end{equation}
with 
\begin{equation}
\mathcal{J} = \left\{\begin{array}{cl} & \exp\left[-(\frac{\tau_{\textrm{dC}}}{\tau_{X}})^{\frac{5}{4}}\right] \hspace{0.51cm}\textrm{for }z<2\times 10^6\\
&\\
& 2.082\left(    \frac{\tau_{\rm dC}}{\tau_{X}}    \right)^{\frac{10}{18}}
\exp\left[-1.988\left( \frac{\tau_{\rm dC}}{\tau_{X}}    \right)^{\frac{10}{18}}\right]\,, \\
& \hspace{4cm} \textrm{otherwise},\end{array}\right.
\end{equation}
where $\tau_{\textrm{dC}} = 1.46 \times 10^8\,(T_0/2.7\textrm{ K})^{-12/5}(\Omega_bh^2)^{4/5}(1-Y_p/2)^{4/5}$ is the ``double Compton'' interaction time in terms of the current CMB temperature $T_0$, with $Y_p\simeq 0.25$ the primordial mass fraction of $^4$He.
We use the limit given by COBE on the chemical potential: $|\mu| \leq 9\times10^{-5}$~\cite{Fixsen:1996nj}, but we will also show the sensitivity that should characterize the future  experiment {\sf PIXIE}, of the order of $\mu \gtrsim 5\times10^{-8}$, at 1-$\sigma$~\cite{Kogut:2011xw}.

\section{Results}\label{results}
One of the most peculiar features of the spectral nonuniversality of photons injected below the pair production threshold is
that the final outcome reflects the energy distribution of the injected photons {\it with respect to} the shape of the relevant
photodisintegration cross sections, shown in Fig.~\ref{fig:Cross_Sections}. This motivated us to choose in the following for each nuclide, the results for two representative examples of monochromatic injection: one close to the resonant peak and another one well after it. The markedly different outcomes obtained in the two cases should thus convincingly argue that constraints of actual models are going to be determined not only  by the decay time and the overall energy injected but also by the energy range at which  the bulk of the photons lies.
 
\subsection{Constraints from ${}^4\textrm{He}$}

\begin{figure}[!h]
\centering
\includegraphics[width=0.49\textwidth]{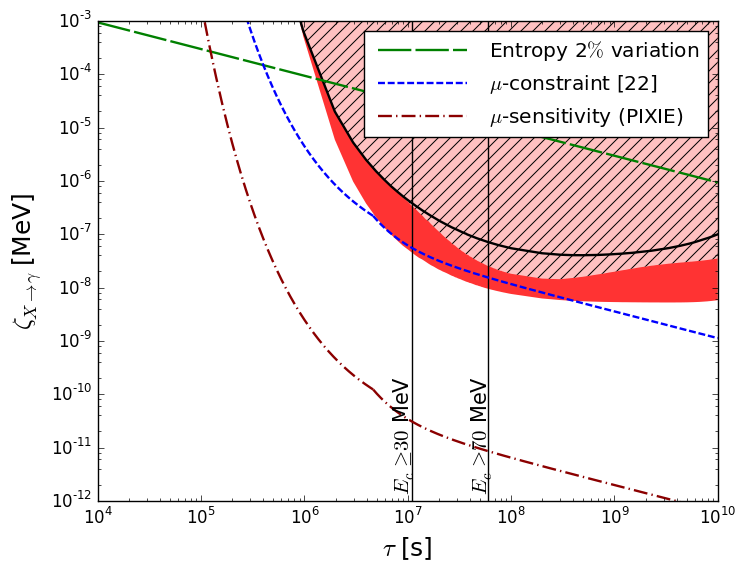}
\caption{\label{fig:Helium} Constraints from $^4$He depletion in the standard case (black line) and for a non-universal spectrum with $E_0$ = 30 MeV (dark shaded red) and $E_0$=70 MeV (hatched/light shaded red). We also show the sensitivity to the entropy variation constraint (green dashed line),  current constraints from CMB spectral distortions (excluded above the short-dashed blue line), and the sensitivity reach of the future mission {\sf PIXIE}~\cite{Kogut:2011xw} (above the red dot-dashed line).}
\end{figure}

The simplest situation is certainly the one concerning ${}^4\textrm{He}$: being the only abundant nucleus subject
to photodisintegration, its nonthermal e.m. production is irrelevant, and one only has to 
care about its destruction; i.e. only the term proportional to $Y_A$ at the rhs of Eq.~(\ref{eq:AbondNy}) is important. 
The results obtained by using a monochromatic injection at 70 MeV (hatched/light shaded red), at 30 MeV (dark shaded red), and the universal
spectrum are shown in Fig.~\ref{fig:Helium}. 
The vertical lines indicate the time at which the threshold energy for pair production $\epsilon_c$ starts exceeding the corresponding injected energy. 
One might naively expect that this is the time at which the constraints obtained from the incorrect use of the universal spectrum  start to deviate from the actual ones. 
However, when taking into account the fact that the decay is not instantaneous, it turns out that constraints already start to deviate at $\sim \tau_X/5$, and  the closer  to the post-threshold cross section resonance we inject energy, the earlier deviations appear.

Although the BBN bounds coming from excessive depletion are typically (but not always!) weaker than CMB distortion bounds, also reported in the figure, note that in both cases bounds can differ from the ones derived with the universal spectrum by a large factor, up to an order of magnitude if the energy of the photons is around the peak of the photodissociation cross section. For higher-injected energies, they tend to become closer to the universal spectrum constraints, as it should. In fact one can envisage
fine-tuned situations in which they become slightly weaker, albeit this conclusion does depend on the extrapolation of the photodisintegration cross sections, of which the reliability at high energy has never been quantitatively assessed in the context of BBN applications.

\subsection{Constraints from ${}^2\textrm{H}$}
For deuterium, the situation is more complicated because several regimes are present.
At low $\tau_X$, $\epsilon_c$ is below ${}^4\textrm{He}$ photodissociation threshold (and in some cases also below  $A=3$ nuclei photodissociation thresholds, which are, however, less relevant).
Hence, only constraints from overdestructions are present.
At high $\tau_X$, however, what dominates is the overproduction from ${}^4\textrm{He}$ destruction.

\begin{figure}[!t]
\centering
\includegraphics[width=0.49\textwidth]{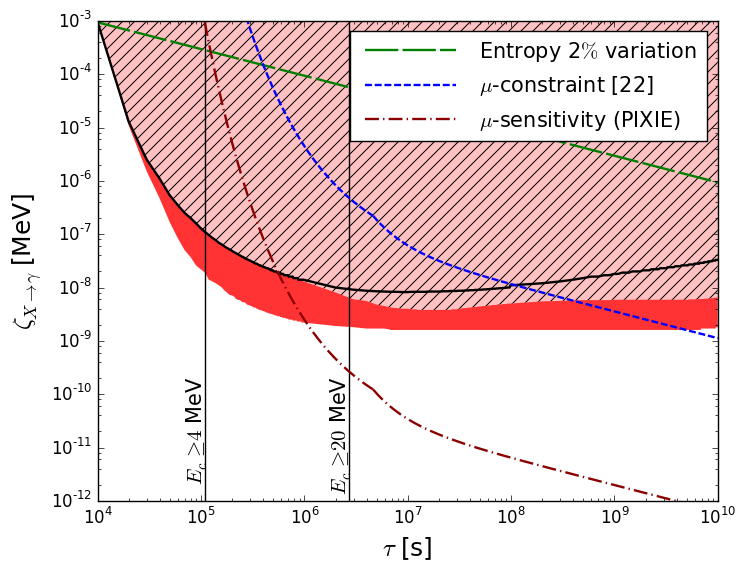}
\caption{\label{fig:Deuterium_Without_Production} Constraints from  deuterium depletion in the standard case, with $E_0 \lesssim 20$ MeV (black solid line) and for a nonuniversal spectrum with $E_0$ = 4 MeV (dark shaded red) and $E_0$=20 MeV (hatched/light shaded red). All other constraints/sensitivities shown as in Fig.~\ref{fig:Helium}.}
\end{figure}

Figure~\ref{fig:Deuterium_Without_Production} shows the illustrative case where production channels are turned off:
this is exact for  $E_0\lesssim 8\,$MeV, but a good approximation till $E_0\lesssim 20\,$MeV.
Note the qualitative similarity to the  ${}^4\textrm{He}$ case, apart for the modifications due to the different features
of the respective cross sections.

\begin{figure}[!t]
\centering
\includegraphics[width=0.49\textwidth]{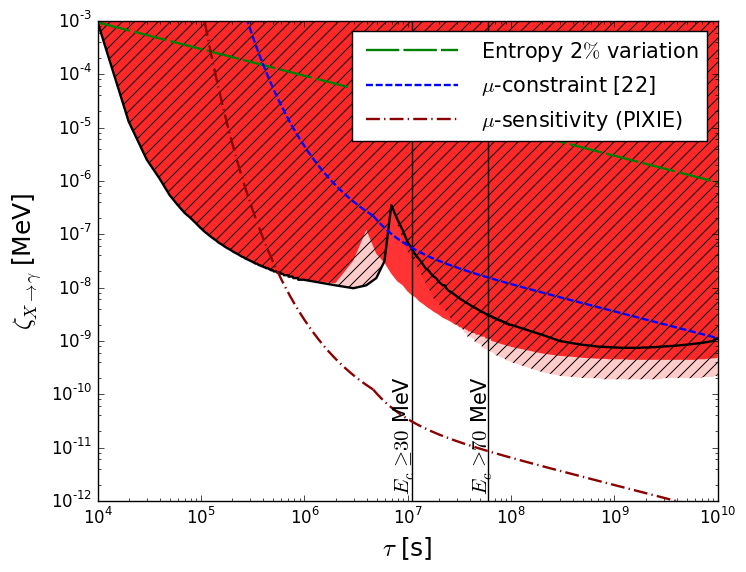}
\caption{\label{fig:Deuterium_With_Production} Constraints from  Deuterium depletion and production in the standard case, with $E_0 \gtrsim 20$ MeV (black solid line)  and for a nonuniversal spectrum with $E_0$ = 30 MeV (dark shaded red) and $E_0$=70 MeV  (hatched/light shaded red). All other constraints/sensitivities shown as in Fig.~\ref{fig:Helium}.}
\end{figure}

Whenever  production channels from ${}^4\textrm{He}$ are open,  which requires $E_0\gtrsim 20\,$MeV,  the constraints are significantly stronger at large $\tau_X$, as shown in  Fig.~\ref{fig:Deuterium_With_Production}. Once again, a violation of universality (and a sensitivity to the energy dependence of the cross section) is clearly manifest by the two cases shown, $E_0= 30\,$MeV
and $E_0=70\,$MeV.


\begin{figure}
\centering
\includegraphics[width=0.49\textwidth]{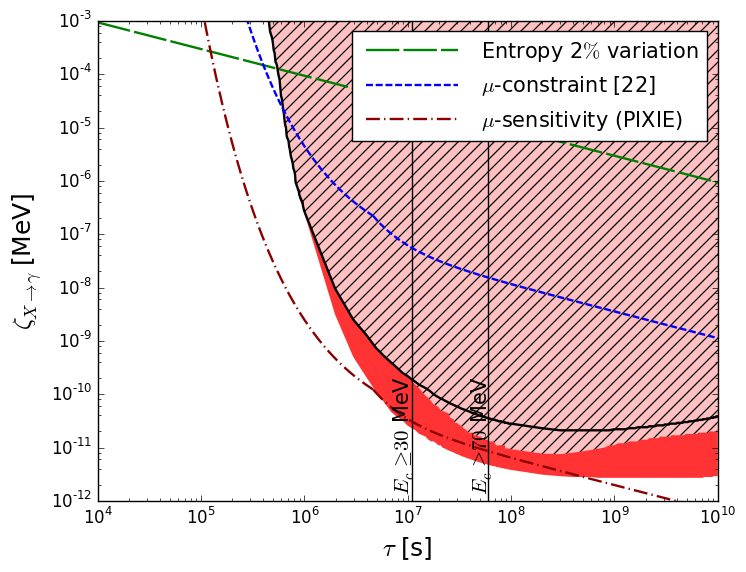}
\caption{\label{fig:Helium3_Non_Universal} Constraints from the $^3$He production in the standard case (black line) and for a nonuniversal spectrum with $E_0$ = 30 MeV (dark shaded red) and $E_0$=70 MeV (hatched/light shaded red). All other constraints/sensitivities shown as in Fig.~\ref{fig:Helium}.}
\end{figure}
\begin{figure}
\centering
\includegraphics[width=0.49\textwidth]{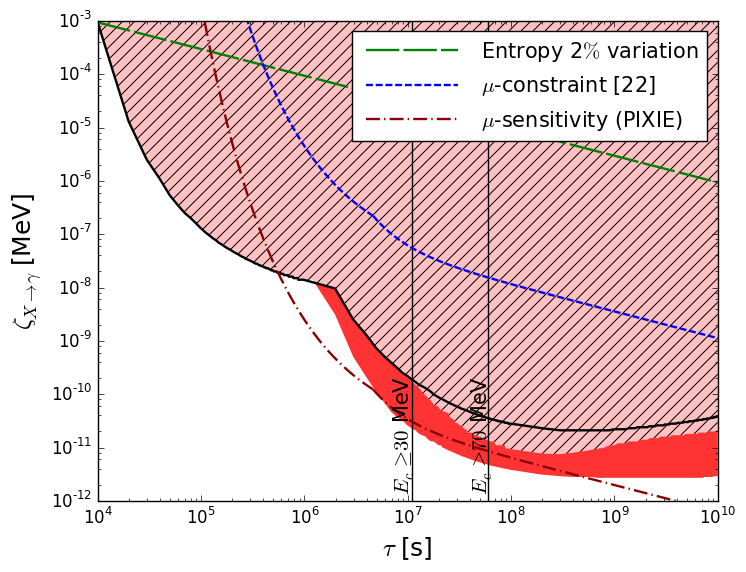}
\caption{\label{fig:Best_Constraints} Global BBN best constraints in the standard case (black line) and for a nonuniversal spectrum with $E_0$ = 30 MeV (dark shaded red) and $E_0$=70 MeV (hatched/light shaded red). All other constraints/sensitivities shown as in Fig.~\ref{fig:Helium}. }
\end{figure}

It is also worth noting that for deuterium the constraints are significantly stronger than the ones coming from CMB spectral distortions.  By improving the sensitivity to $\mu$ down to $\mu \gtrsim 5\times 10^{-8}$, the sensitivity expected by the future mission {\sf PIXIE}~\cite{Kogut:2011xw} (shown by the red, dot-dashed curve) would greatly strengthen these constraints, with the exception of very small lifetimes where deuterium overdestruction would still provide the dominant bounds.

\subsection{Constraints from ${}^3\textrm{He}$ }
 First of all, a premise is necessary: there are in fact two nuclei with $A=3$,  ${}^3\textrm{He}$, and  ${}^3\textrm{H}$, the latter being unstable to beta decay into  ${}^3\textrm{He}$ with a half-time of over 12 years, or about  $4\times 10^8\,$s. Practically, however, for the purposes of the constraints discussed here, one can sum the equations for ${}^3\textrm{He}$ and ${}^3\textrm{H}$ and treat them as a single effective nucleus with $A=3$. The reason is twofold: first, we only require ${}^3\textrm{He}$  not to be overproduced with respect to the observational upper limit. Hence, the key reactions are the {\it production} channels by single nucleon photodisintegration from ${}^4\textrm{He}$, which are only open above 20 MeV, rather than the destruction ones. Second, ${}^3\textrm{He}$ and  ${}^3\textrm{H}$ are ``mirror nuclei'' under the isospin symmetry $n\leftrightarrow p$, and their nuclear properties are in fact very similar: the corresponding thresholds in nuclear cross sections, for instance, only differ by some 0.8 MeV (compare the two curves in Fig.~\ref{fig:Cross_Sections}.)
From Fig.~\ref{fig:Helium3_Non_Universal}, where we report our results, it is clear that the photodisintegration cross section for single nucleon emission from ${}^4\textrm{He}$, when open, is so important that very stringent nucleosynthesis constraints follow.
In fact, they are much stronger than the current ones coming from CMB spectral distortions, although future {\sf PIXIE} sensitivity might improve over them over most of the parameter space. 

Notice the importance of
the nonuniversality: the two cases with 30 or 70 MeV monochromatic injections lead to significantly different constraints,
and in both cases depart from the ``universal spectrum'' ones. 

\section{Conclusions}\label{concl}
We have  argued that the universality of the photon spectral shape in electromagnetic cascades has often been
used in cosmology even beyond its regime of applicability. When the energy of the injected
photons falls below the pair production threshold, i.e.  approximately when $E_\gamma\lesssim m_e^2/(22\,T)\sim 10\, T_{\rm keV}^{-1}\,$MeV, the universal form breaks down. In Ref.~\cite{PaperI}, we showed how this could potentially open the possibility
of a purely electromagnetic decay solution to the so-called lithium problem. In this article, we 
showed how important the modifications to the photon spectrum in this regime are for the constraints from nonthermal
BBN. This required the numerical solution of the relevant Boltzmann equations, which we attacked by an iterative scheme.

The constraints we obtained, for illustrative cases of monochromatic energy injection at different epochs, 
are often much stronger than the ones presented in the literature (up to an order of magnitude), notably when the
injected photon energy falls close to the peak of the photodisintegration cross section of the relevant nucleus.
In fact, the breaking of the nonuniversality is nontrivial and is essentially controlled by the energy behavior of the cross sections: in the universal limit, most of the photons
lie at relatively low-energies, so that the cross-section behaviour at the resonance just above threshold is what matters
the most. In the actual treatment, the photons may be also sensitive to the high-energy tail of the process. Future studies aiming at assessing the nuclear physics uncertainties affecting these types of bounds  would benefit from this insight. It cannot be excluded that in some cases constraints {\it weaken} a bit with respect to what 
is considered in the literature.

We also compared BBN bounds with constraints coming from CMB spectral distortions. A summary plot of the ``best constraints'' is reported
in Fig.~\ref{fig:Best_Constraints}, for two choices of the monochromatic photon energy. We concluded that BBN limits are improving over {\it current} constraints from COBE via the requirement not to underproduce ${}^2\textrm{H}$ (at low injection
lifetime $\tau_X$), or not to overproduce  ${}^3\textrm{He}$ (at high  $\tau_X$), while ${}^4\textrm{He}$ is never competitive.
The bounds from a future CMB spectral probe such as {\sf PIXIE} would not only greatly improve current CMB constraints 
but would also reach the level of current constraints from ${}^3\textrm{He}$ (often improving over them) allowing for an independent consistency check.
This is reassuring, since the cosmological reliability of ${}^3\textrm{He}$ constraints does stand on some astrophysical assumptions. Below $\tau_X\sim 5\times 10^5\,$s, however, ${}^2\textrm{H}$ constraints would probably remain the most stringent ones for a long time to come.  Fortunately they are i) quite robust, relying on the single, well-known cross section ${}^2\textrm{H}(\gamma,n)p$, and ii) easy to compute, since no coupled network of equations needs to be solved, the problem reducing to the numerical evaluation of a single integral (the same situation leading to Eq.~(6) in Ref.~\cite{PaperI}.)
This is also the region where constraints coming from hadronic decay modes (not revisited here) are quite stringent. A synergy between BBN and CMB is thus going to be necessary for this kind of physics even in the decades to come. 

In conclusion, our work suggests  that models in the literature that fulfilled the BBN constraints with less than an order of magnitude margin should perhaps be reconsidered. In particular, those characterized by soft gamma-ray emissions and/or at relatively late times should have been more prone to incorrect conclusions about their viability.  
Our study also suggests that actual bounds should be derived via a case-by-case analysis. Finally, we provided further arguments supporting the usefulness of an improved constraint from CMB spectral distortion of the $\mu$ type, since it would not manifest the unexpected sensitivity to the shape of the energy injection that we have uncovered.

\acknowledgments
Support by the Labex Grant ENIGMASS is acknowledged.


\end{document}